\documentclass[a4paper,11pt]{article}
\usepackage{pos}

\usepackage{slashed,physics}
\usepackage{dsfont}
\renewcommand{\logo}{\relax}

\newcommand{\ba}{\begin{array}}
\newcommand{\ea}{\end{array}}

\newcommand\aNLO{{\sc\small MadGraph5\_aMC@NLO}}
\newcommand\WZ{\sc\small Whizard}

\title{Higgs-muon interactions at a multi-TeV muon collider}
\author*[a,b]{Yang Ma}
\affiliation[a]{INFN, Sezione di Bologna, via Irnerio 46, 40126 Bologna, Italy}
\affiliation[b]{Center for Cosmology, Particle Physics and Phenomenology, Universit\'e catholique de Louvain, B-1348 Louvain-la-Neuve, Belgium}

\author[c]{Eugenia Celada}
\affiliation[c]{Department of Physics and Astronomy, University of Manchester, Oxford Road, Manchester M13 9PL, United Kingdom}

\author[d]{Tao Han}
\affiliation[d]{Pittsburgh Particle Physics, Astrophysics, and Cosmology Center, Department of Physics and Astronomy, University of Pittsburgh, 3941 O'Hara St
Pittsburgh, PA 15260, USA}

\author[e]{Wolfgang Kilian}
\author[e]{Nils Kreher}
\affiliation[e]{Center for Particle Physics Siegen, University of Siegen, Walter-Flex-Str. 3, 57072 Siegen, Germany}

\author[a,b,f]{Fabio Maltoni}
\author[a]{Davide Pagani}
\affiliation[f]{Dipartimento di Fisica e Astronomia, Universit\`a di Bologna, via Irnerio 46, 40126 Bologna, Italy}

\author[g]{J\"urgen Reuter}
\affiliation[g]{Deutsches Elektronen-Synchrotron DESY, Notkestr. 85, 22607 Hamburg,
Germany}

\author[e]{Tobias Striegl}
\author[h]{Keping Xie}
\affiliation[h]{Department of Physics and Astronomy, Michigan State University, 567 Wilson Road, East Lansing, MI 48824, USA}

\emailAdd{yang.ma@bo.infn.it}
\emailAdd{eugenia.celada@postgrad.manchester.ac.uk}
\emailAdd{than@pitt.edu}
\emailAdd{kilian@physik.uni-siegen.de}
\emailAdd{nils.kreher@uni-siegen.de}
\emailAdd{fabio.maltoni@cern.ch}
\emailAdd{davide.pagani@bo.infn.it}
\emailAdd{juergen.reuter@desy.de}
\emailAdd{tobias.striegl@physik.uni-siegen.de}
\emailAdd{xiekepi1@msu.edu}

\abstract{
  We establish a simple yet general parameterization of Higgs-muon interactions within the effective field theory frameworks, including both the Higgs Effective Field Theory (HEFT) and the Standard Model Effective Field Theory (SMEFT). We investigate the potential of a muon collider, operating at center-of-mass energies of 3 and 10 TeV, to probe Higgs-muon interactions. All possible processes involving the direct production of multiple electroweak bosons ($W$, $Z$, and $H$) with up to five final-state particles are considered. Our findings indicate that a muon collider can achieve greater sensitivity than the high-luminosity LHC, especially considering the independence of the Higgs decay branching fraction to muons. Notably, a 10 TeV muon collider offers exceptional sensitivity to muon-Higgs interactions, surpassing the 3 TeV option. In particular, searches based on multi-Higgs production prove highly effective for probing these couplings.
}

\FullConference{42nd International Conference on High Energy Physics (ICHEP2024)\\
18-24 July 2024\\
Prague, Czech Republic\\}

\begin{document}
\maketitle

\section{Introduction}
\label{sec:intro}
\noindent
As the Higgs boson is believed to be special in the Standard Model (SM) of particle physics and also a portal to possible new physics beyond the Standard Model (BSM), precise measurements on its interactions with other SM particles are critical. 
While the Yukawa couplings of the third-generation fermions ($t$, $b$, and $\tau$) have been measured to be consistent with the SM predictions, probing the interactions of the second-generation fermions with the Higgs boson remains a priority.
Evidence for the Higgs-muon coupling has already emerged from Large Hadron Collider (LHC) measurements of the $H \to \mu^+ \mu^-$ decay channel \cite{CMS:2020xwi,ATLAS:2020fzp}, and the High-Luminosity LHC (HL-LHC) is expected to improve this measurement with higher precision.
However, these measurements depend on the assumption of the Higgs boson's total decay to be the SM value. 
% To achieve a model-independent measurement of the Higgs-muon interaction, exploring options beyond the LHC is necessary. 
Meanwhile, as a promising next-generation lepton collider, a multi-TeV muon collider~\cite{Black:2022cth,Accettura:2023ked} combines the advantages of high-energy hadron colliders and $e^+e^-$ colliders~\cite{Costantini:2020stv,Han:2020uid,Han:2021kes}. In our previous work~\cite{Han:2021lnp,Reuter:2022zuv}, we demonstrated that multi-boson production offers a valuable opportunity to measure the $H{\bar \mu} \mu$ vertex. In this proceeding, we summarize our recent study of the comprehensive parameterization within the effective field theory (EFT) framework and include multi-Higgs production processes in the analysis ~\cite{Celada:2023oji}. 

\section{Model Parameterization}
To provide a simple yet general parameterization of Higgs-muon interactions, we introduce the form factors $\alpha_i$ and $\beta_j$ to describe the couplings of $\bar{\mu}\mu H^i$ and $H^j$ interactions, respectively. 
In the unitary gauge, the Lagrangian is expressed as: 
\begin{eqnarray}
  \mathcal{L}\supset  -\frac{m_H^2}{2}H^2 - m_\mu \bar{\mu}\mu - \sum_{n=3}^\infty \beta_n \frac{\lambda}{v^{n-4}} H^n - \sum_{n=1}^\infty \alpha_n \frac{m_\mu}{v^n}\bar{\mu}\mu H^n. \label{eq:LextendedK}
\end{eqnarray}
The above Lagrangian aligns with the Higgs Effective Field Theory (HEFT), where the relations $y_{\mu,n} = \sqrt{2} m_\mu \alpha_n / v$ and $f_{V,n} = \beta_n \lambda$ hold~\cite{Han:2021lnp}.
By adopting $\{\alpha_1=1,\alpha_{n>1}\}=\{1,0\}$ for muon-Higgs interactions and $\{\beta_3,\beta_4,\beta_{n>4}\}=\{1,1/4,0\}$ for Higgs self-interactions, the Lagrangian in Eq.~\eqref{eq:LextendedK} reduces to the SM one. 
Focusing on the Yukawa sector, the form factors $\alpha_n$ can be related to the Standard Model Effective Field Theory (SMEFT) parameters $c_{\ell \varphi}^{(n)}$ as follows:
\begin{equation}\label{eq:map}
\begin{aligned}
    &\alpha_1= 1+ \frac{v^3}{\sqrt 2 m_\mu} \frac{c_{\ell \varphi}^{(6)}}{\Lambda^2}+ \frac{v^5}{\sqrt 2 m_\mu} \frac{c_{\ell \varphi}^{(8)}}{\Lambda^4}+ \frac{3v^7}{4\sqrt 2 m_\mu} \frac{c_{\ell \varphi}^{(10)}}{\Lambda^6}\,, \\
    &\alpha_2=  \frac{3v^3}{2\sqrt 2 m_\mu} \frac{c_{\ell \varphi}^{(6)}}{\Lambda^2} +\frac{5 v^5}{2\sqrt 2 m_\mu} \frac{c_{\ell \varphi}^{(8)}}{\Lambda^4}+\frac{21 v^7}{8\sqrt 2 m_\mu} \frac{c_{\ell \varphi}^{(10)}}{\Lambda^6}\,,  \\
  &\alpha_3 =  \frac{v^3}{2\sqrt 2 m_\mu} \frac{c_{\ell \varphi}^{(6)}}{\Lambda^2} +\frac{ 5 v^5}{ 2\sqrt 2m_\mu} \frac{c_{\ell \varphi}^{(8)}}{\Lambda^4}+\frac{35v^7}{8\sqrt 2 m_\mu}\frac{c_{\ell\varphi}^{(10)}}{\Lambda^6}\,,  \\
  &\alpha_4 =  \frac{ 5 v^5}{ 4\sqrt 2m_\mu} \frac{c_{\ell \varphi}^{(8)}}{\Lambda^4}+\frac{35v^7}{8\sqrt 2 m_\mu}\frac{c_{\ell\varphi}^{(10)}}{\Lambda^6}\,, \alpha_5 = \frac{v^5}{ 4\sqrt 2m_\mu} \frac{c_{\ell \varphi}^{(8)}}{\Lambda^4}+\frac{21v^7}{8\sqrt 2 m_\mu}\frac{c_{\ell\varphi}^{(10)}}{\Lambda^6}\,,
  \phantom{+\frac{35v^7}{\sqrt 2 m_\mu}\frac{c_{\ell\varphi}^{(10)}}{\Lambda^6}}
  \end{aligned}
  \end{equation}
  where we stop up to dimension ten for the SMEFT operators. Specially, in the dim-6 SMEFT scenario, the relations $\Delta\alpha_1\equiv \alpha_1-1=\frac{2}{3}\alpha_2=2\alpha_3$ and $\alpha_4=\alpha_5=0$ hold.

\section{Phenomenology}
We consider possible high-energy muon colliders with collision energies 3~TeV (10~TeV) with luminosities $1~{\rm ab}^{-1}$ ($10~{\rm ab}^{-1}$)~\cite{Black:2022cth,Accettura:2023ked}, respectively. The signal is defined as the set of direct muon-annihilation processes into multiple bosons $\mu^+\mu^-\to mV+nH$ up to $m+n=5$, where $V=W,Z$ and $H$ is the Higgs boson.
As suggested in Ref.~\cite{Han:2021lnp,Reuter:2022zuv}, we apply the kinematic cuts on the final-state bosons
\begin{eqnarray}
  \theta_{iB} > 10^\circ, \quad \Delta R_{BB}>0.4, \quad M_F > 0.8 \sqrt {s},
\end{eqnarray}
where $\theta_{iB}$ is the smallest angle between any final-state boson $B$ ($B=H,W,Z$) and the beam axis, $\Delta R_{BB}=\sqrt{\Delta\eta^2+\Delta \phi^2}$ is the separation distance between any two bosons, and $M_F$ is the invariant mass of all final-state bosons. The invariant mass cut is sufficient to suppress the vector boson fusion (VBF) backgrounds and to reconcile the initial state radiation (ISR). The numerical results are obtained using the general-purpose generators {\aNLO}~\cite{Alwall:2014hca, Frederix:2018nkq} and {\WZ}~\cite{Kilian:2007gr}.
Other relevant effects from the experimental simulation ({\it e.g.}, vector-boson tagging efficiency) and theorical consideration ({\it e.g.}, PDF effects~\cite{Han:2020uid,Han:2021kes} or NLO EW corrections~\cite{Bredt:2022dmm,Ma:2024ayr}) are reserved for a future study, which is not expected to change our conclusions dramatically.

% Comment out for length limit
% Assuming uncertainties dominated by statistics, the experimental sensitivity can be obtained by the significance ${\mathcal S}$, defined as
% \begin{eqnarray}
% \label{eq:sen}
% {\mathcal S} =  \sqrt{2(S+B)\log(1+\frac{S}{B})-2S}\,,
% \end{eqnarray}
% where $S$ and $B$ denote the signal and background event numbers, ${\mathcal S} =2~(3)$ corresponds to the $2~(3)~\sigma$ exclusion limit, \emph{i.e.}, the 95\% (99\%) confidence level (CL). In some special cases, the background is negligible, i.e. $B\ll 1$, we take  $S=3$ for $95\%$ CL. 

\subsection{Multi-Higgs production}
\begin{figure}[htb]
      \includegraphics[width=0.245\textwidth]{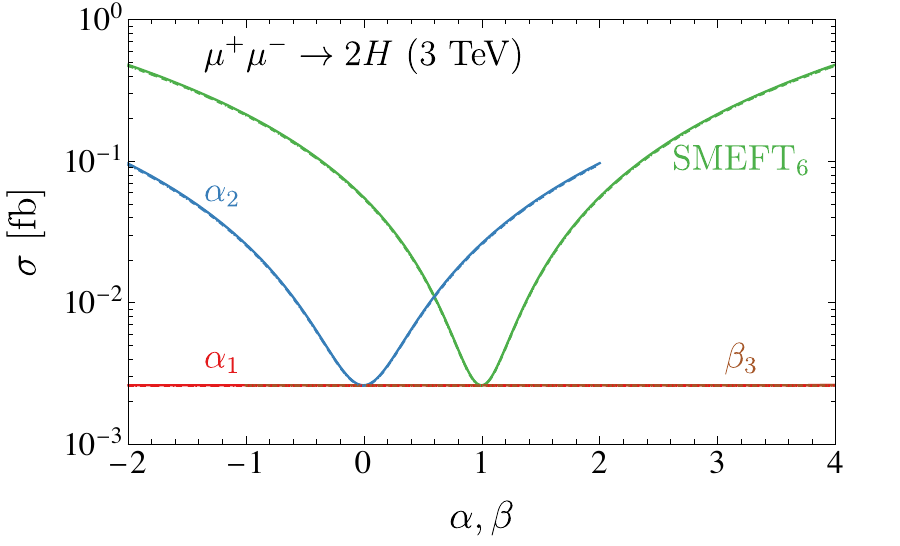}
      \includegraphics[width=0.245\textwidth]{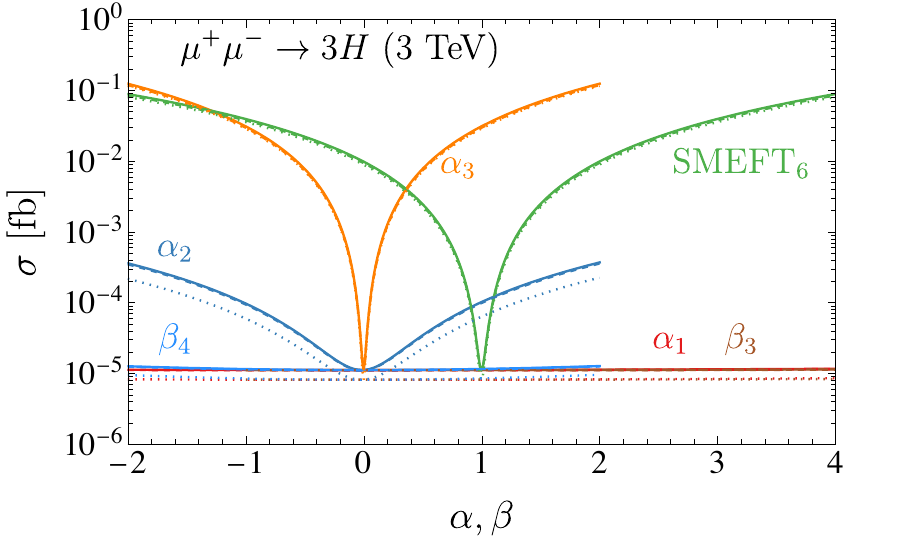}
      \includegraphics[width=0.245\textwidth]{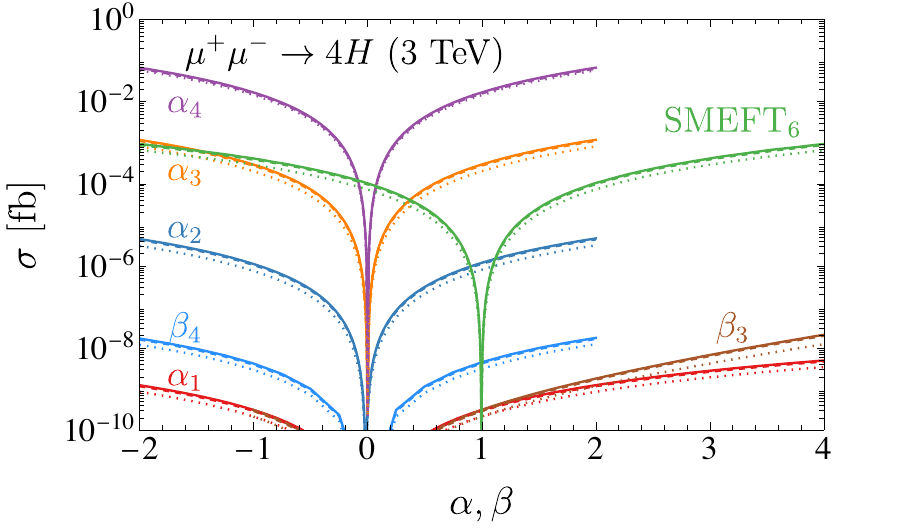}
      \includegraphics[width=0.245\textwidth]{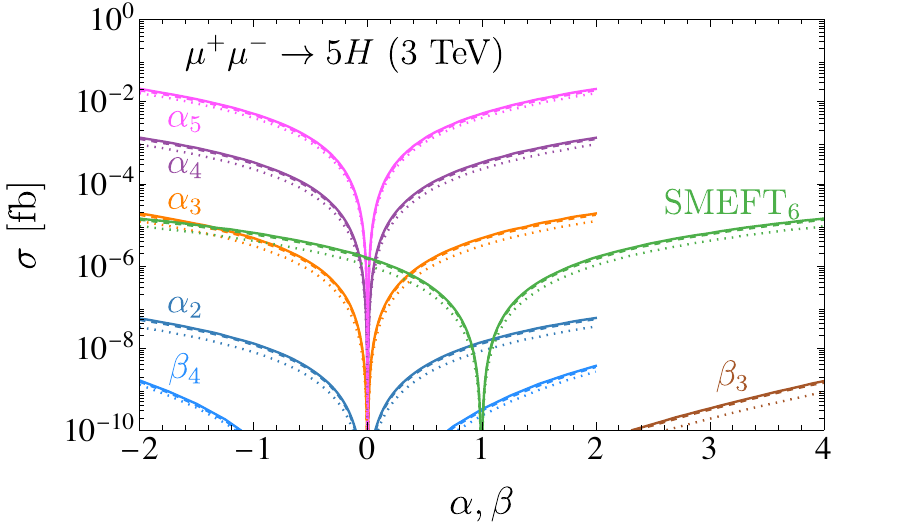}
      \caption{The cross sections of $\mu^+\mu^- \to nH$ as functions of the parameters $\alpha_i$ and $\beta_j$ at 3~TeV. The green curves are for $\alpha_1=1+\Delta \alpha_1$ in the dim-6 SMEFT scenario. 
      Solid lines refer to the configuration with no cuts, dashed lines to the case with $|\theta_{iB}|>10^\circ$ cuts applied and dotted lines to the case with all cuts applied. Adapted from Ref.~\cite{Celada:2023oji}.}
      \label{fig:mmnhkappa3}
\end{figure}

Due to the smallness of the SM $\bar{\mu}\mu H$ coupling, the tree-level contribution to the multi-Higgs production cross section $\sigma_{\rm SM}^{\rm LO}$ is highly suppressed and the dominant contribution originates from the square of one-loop diagrams ($\sigma_{\rm SM}^{\rm loop}$).
%
% Comment out for length limit
% Including possible BSM effects, the multi-Higgs production cross sections can be written as
% \begin{eqnarray}
% \sigma_{\rm BSM}(\mu^+\mu^-\to nH)=\sigma_{\rm SM}^{\rm loop} + \sigma_n (\alpha_n^2) +  \sigma_{\rm sub}(\alpha_n,\alpha_{n-1}, \dots
% ,\alpha_1) \, ,
% \end{eqnarray}
% where $\sigma_n (\alpha_n^2)$ represents the contribution from the $\bar\mu\mu H^n$ vertex alone and $\sigma_{\rm sub}(\alpha_n,\alpha_{n-1}, \dots ,\alpha_1)$ contains all the possible $\alpha_i$ vertices with $i\le n$. 
%
In practice, we have considered $\sigma_{\rm SM}^{\rm loop}$ for only the di-Higgs and tri-Higgs production, and the beyond is expect to be suppressed as well. We present the sensitivities of $\sigma_{\rm BSM}$ on the $\alpha_i$ and $\beta_j$ parameters in Figure~\ref{fig:mmnhkappa3}, showing that the multi-Higgs production processes enable unique measurements on the $\bar\mu\mu H^n$ vertices. The upper bounds of $\alpha_n$ and the corresponding signal strength are summarized in Table~\ref{tab:mmnhsig}. Keeping only the dim-6 operators, we obtain a translation
\begin{eqnarray}
|\Delta \alpha_1|\lesssim 0.3 ~~\Longleftrightarrow~~\left|c^{(6)}_{\ell\varphi}/\Lambda^2\right| &\lesssim& 3\times 10^{-9} \,{\rm GeV}^{-2} \qquad{\rm at~3~TeV}\,, \label{eq:boundmultiHEFT1a}\\
|\Delta  \alpha_1|\lesssim 0.1 ~~\Longleftrightarrow~~\left|c^{(6)}_{\ell\varphi}/\Lambda^2\right| &\lesssim& 1 \times 10^{-9} \,{\rm GeV}^{-2} \qquad{\rm at~10~TeV}\,, \label{eq:boundmultiHEFT1b}
\end{eqnarray}
via measuring the $2H$ production, while from the measurement of $3H$ production we have 
 \begin{eqnarray}
|\Delta  \alpha_1|\lesssim 0.7 ~~\Longleftrightarrow~~  \left|c^{(6)}_{\ell\varphi}/\Lambda^2\right| &\lesssim& 7\times 10^{-9}\, {\rm GeV}^{-2}   \qquad{\rm at~3~TeV}\,, \label{eq:boundmultiHEFT2a}\\
|\Delta  \alpha_1|\lesssim 0.05 ~~\Longleftrightarrow~~ \left|c^{(6)}_{\ell\varphi}/\Lambda^2\right| &\lesssim& 5\times 10^{-10}\, {\rm GeV}^{-2}   \qquad{\rm at~10~TeV}\,. \label{eq:boundmultiHEFT2b}
 \end{eqnarray}

\begin{table}[htb]
  \centering
  \scalebox{0.8}{
  \begin{tabular}{c|ccc|ccc}
  \hline
  $\sqrt{s}$ & \multicolumn{3}{c|}{3 TeV}  & \multicolumn{3}{c}{10 TeV} \\
  \hline
  $n$   & bound on $|\alpha_n|$   & $S/B$  & $\mathcal{S}$     & bound on $|\alpha_n|$   & $S/B$  & $\mathcal{S}$     \\ \hline
  2 & 0.42 & 1.49 & 2.05  & 0.15 & 1.16  & 2.12 \\
  3 & 0.33  & -- & --  &  $2.6 \cdot 10^{-2} $ & 3.23 & 2.03 \\
  4 & 0.46 & -- & -- & $1.4 \cdot 10^{-2} $ & 1.31 & 2.00 \\
  5 & 0.87 & -- &  -- & $9.0\cdot 10^{-3}$ & 0.757 & 2.03  \\
  \hline
  \end{tabular}
  }
  \caption{Bounds on the signal strength of $\alpha_n$ from the $\mu^+\mu^- \to n H$ processes. For $nH$ ($n\geq 3$) production at the 3 TeV muon collider, the SM background gives $\sim 0$ event and the bound on $|\alpha_n|$ are taken from $S=3$.}\label{tab:mmnhsig}
\end{table} 

% \begin{figure}[!t]
%       \centering
%       \includegraphics[width=0.32\textwidth]{figs/WWHkappa3TeV}
%       \includegraphics[width=0.32\textwidth]{figs/ZHHkappa3TeV}
%       \includegraphics[width=0.32\textwidth]{figs/ZZHkappa3TeV}
%       \includegraphics[width=0.32\textwidth]{figs/WWHkappa10TeV}
%       \includegraphics[width=0.32\textwidth]{figs/ZZHkappa10TeV}
%       \includegraphics[width=0.32\textwidth]{figs/ZHHkappa10TeV}
%       \caption{Same as Figure~\ref{fig:mmnhkappa3} for $WWH$, $ZZH$, $ZHH$ production at 3 and 10 TeV, respectively.}
%       \label{fig:mmvvhkappa}
%     \end{figure}

\subsection{Higgs-associated gauge boson production and multi-gauge boson production}
\begin{figure}[htb]
  \centering
  \includegraphics[width=0.32\textwidth]{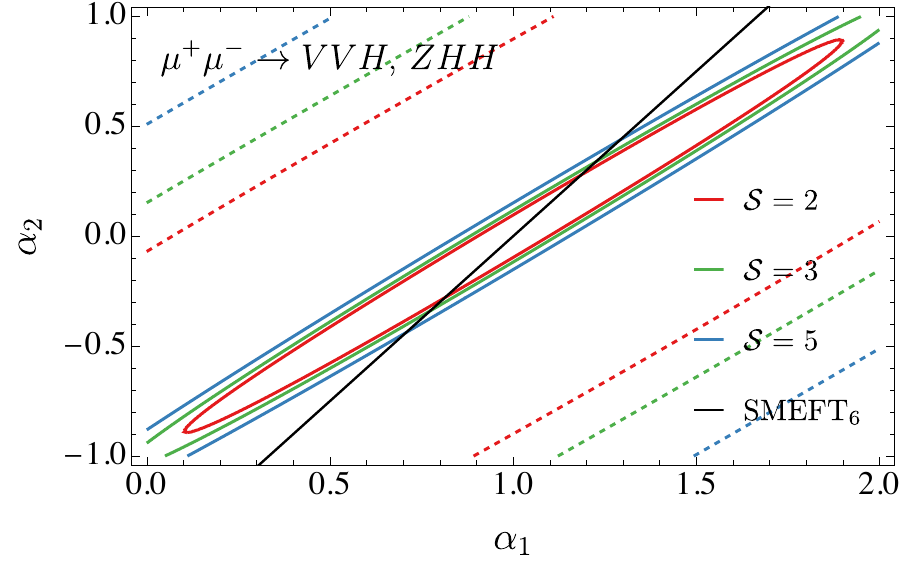}
  \includegraphics[width=0.32\textwidth]{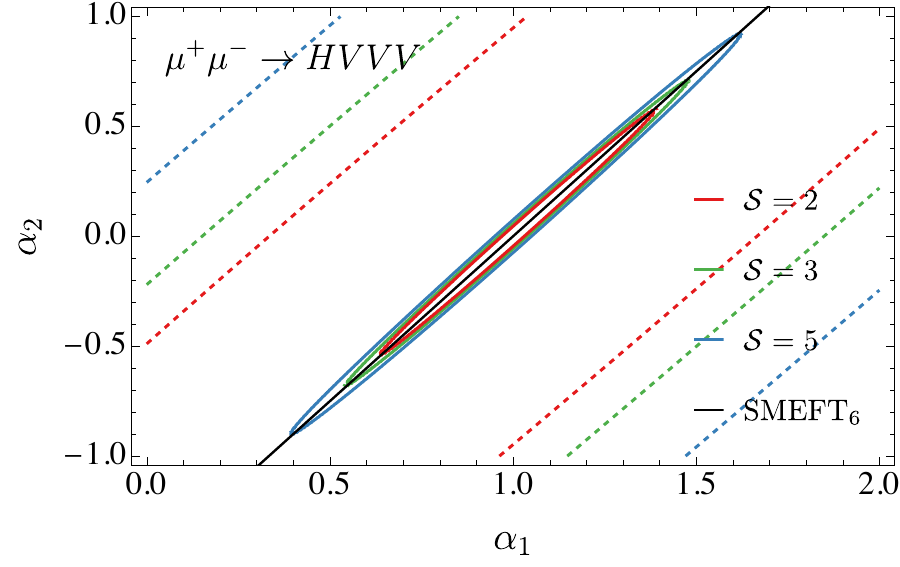}
  \includegraphics[width=0.32\textwidth]{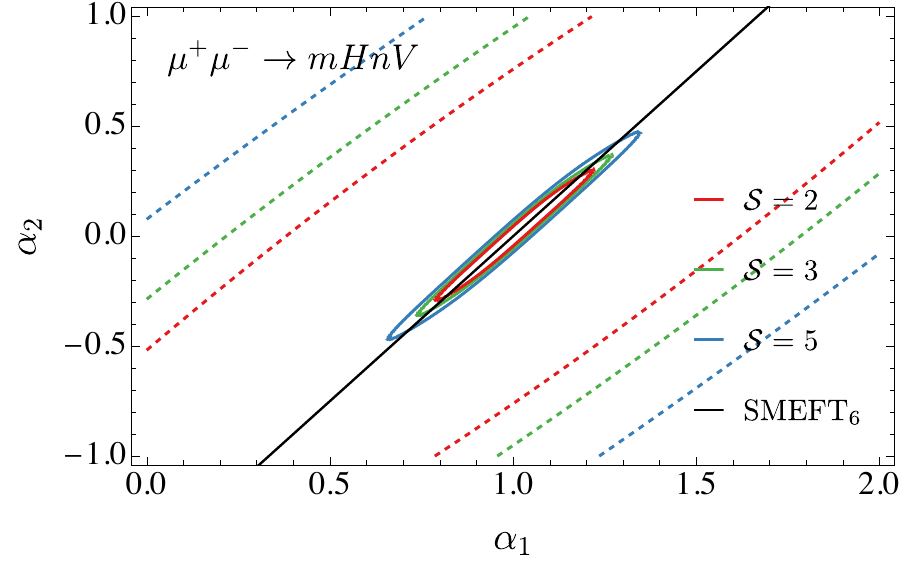}
  \caption{Contour-plots displaying the constraints in the $(\alpha_1,\,\alpha_2)$ plane from Higgs-associated gauge boson production processes for three-boson final states (left) and $\mu^+\mu^- \to V^3 H$ (middle), and the combined signal significance (right) at a 3 TeV muon collider (dashed curves) and a 10 TeV muon collider (solid curves), respectively. The red, green, and blue curves represent the ${\mathcal S}=2,\, 3,\, 5$ significances, respectively. The black solid line corresponds to the dim-6 SMEFT scenario. Adapted from Ref.~\cite{Celada:2023oji}.}
  \label{fig:contour34VH}
\end{figure}

\begin{figure}[!t]
      \centering
      \includegraphics[width=0.32\textwidth]{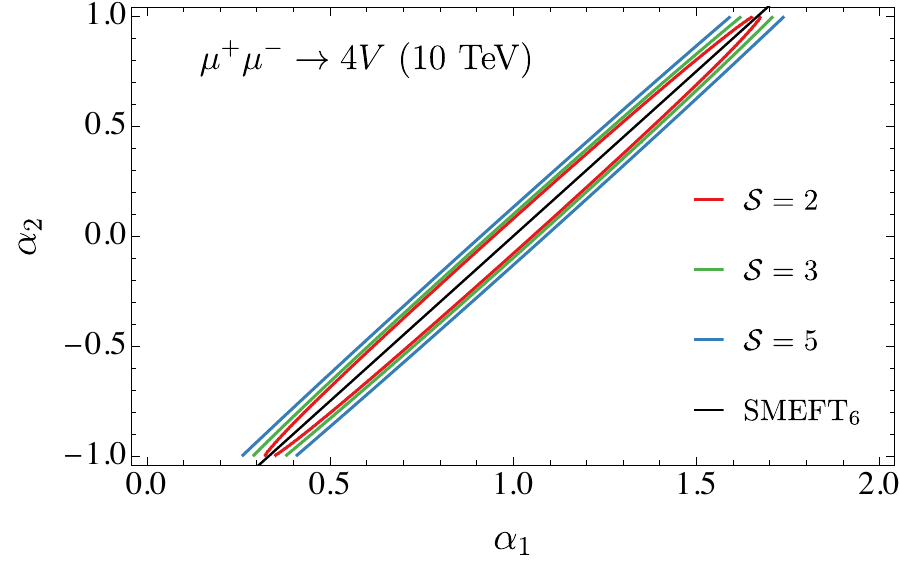}
      \includegraphics[width=0.32\textwidth]{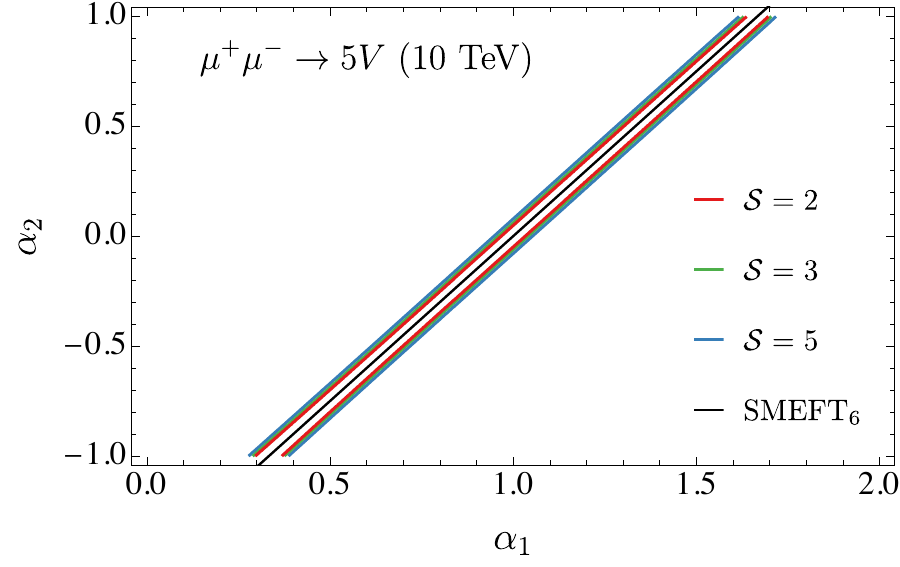}
      \includegraphics[width=0.32\textwidth]{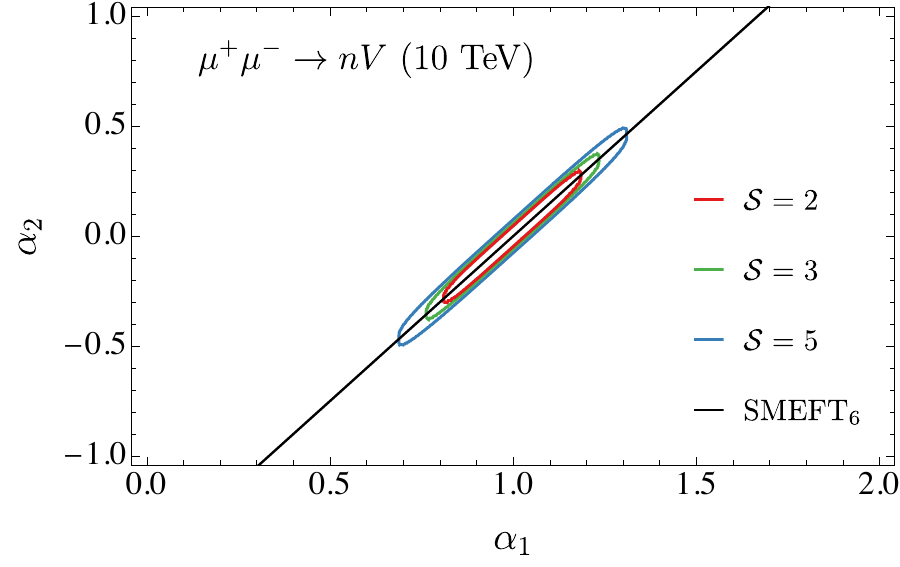}
      \caption{Same as Figure~\ref{fig:contour34VH} for $\mu^+\mu^- \to 4 V$ (left) and $\mu^+\mu^- \to 5 V$ (middle) production at a 10 TeV  muon collider. The right plot shows the combined constraints. Adapted from Ref.~\cite{Celada:2023oji}.}
      \label{fig:contour45V}
    \end{figure}

\begin{figure}[!t]
      \centering
      \includegraphics[width=0.42\textwidth]{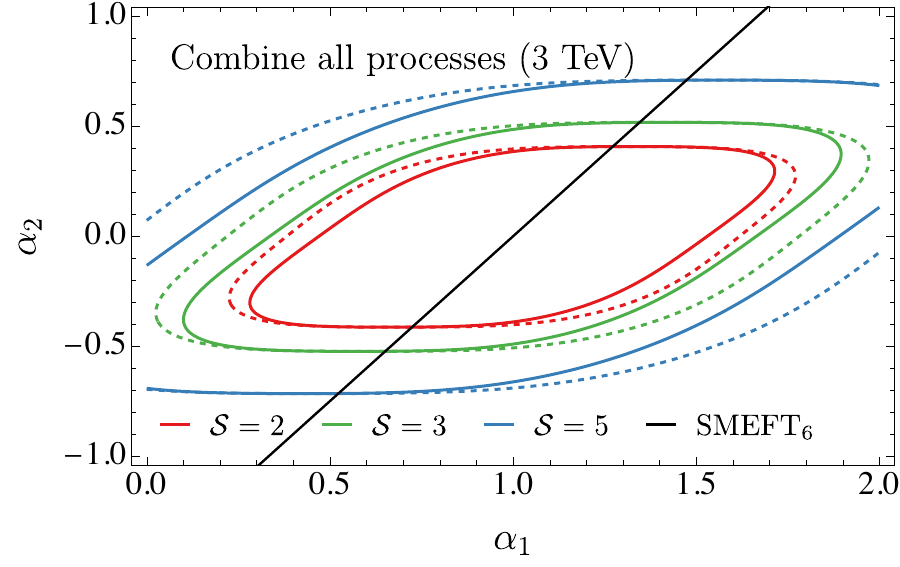}
      \includegraphics[width=0.42\textwidth]{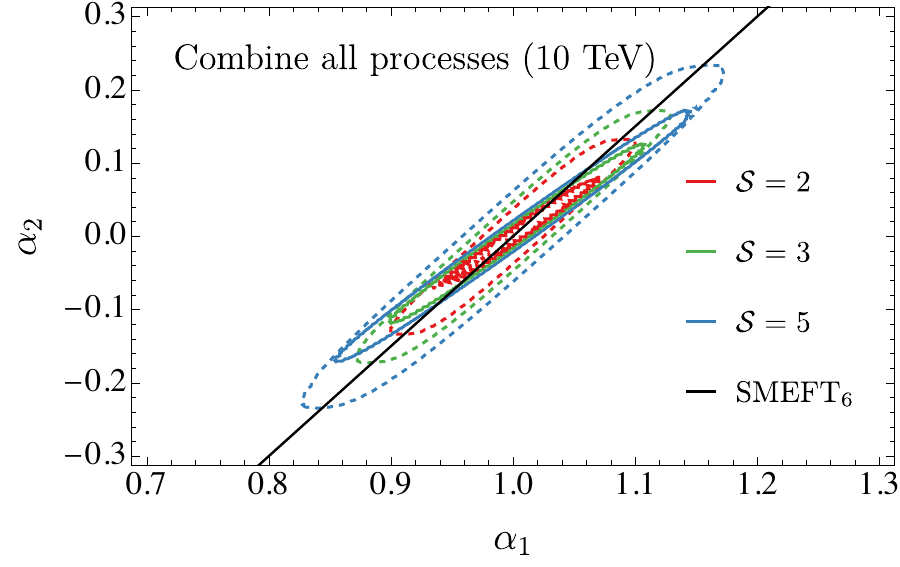}
      \caption{Combined constraints on $(\alpha_1,\,\alpha_2)$ from combining the processes at a 3 TeV muon collider (left) and a 10 TeV muon collider (right), respectively. The dashed curves are for the constraints with no assumptions and the solid curves includes also the processes with assumption $\alpha_3=0$. The red, green, and blue curves represent ${\mathcal S}=2,\, 3,\, 5$ significances, respectively. The black solid line corresponds to the dim-6 SMEFT scenario. Adapted from Ref.~\cite{Celada:2023oji}.}
      \label{fig:contoura1a2all}
    \end{figure}

While the $ZH$ and $3V$ production processes depend solely on $\alpha_1$~\cite{Celada:2023oji}, their sensitivities are too weak to give meaningful constraints. 
The dependence on $\alpha_{n>1}$ emerges in processes with higher multiplicities, summarized as follows.
\begin{itemize}
  \item All the 3-boson final states and $V^3H$ productions are dependent on $\alpha_1$ and $\alpha_2$, with the corresponding measurements probing these two parameters simultaneously, shown in Figure~\ref{fig:contour34VH}. As a reference, we also show the dim-6 SMEFT scenario $ \alpha_1=1+\frac{2}{3}\alpha_2$ as a black solid line.
  \item At a 10~TeV muon collider, the $\mu^+\mu^- \to 4V$ and $\mu^+\mu^- \to 5V$ processes are also sensitive to $\alpha_1$ and $\alpha_2$ as shown in Figure~\ref{fig:contour45V}. Combining these processes could help to improve the constraints on $\alpha_1$ and $\alpha_2$.
  \item Other processes, such as $V^2H^2$, $V^4H$, $ZH^3$, $V^3H^2$ production, are sensitive also on $\alpha_3$. Assuming $\alpha_3=0$, the constraints on $\alpha_1$ and $\alpha_2$ can be further improved. In Figure~\ref{fig:contoura1a2all}, we combine all the processes to show the constraints on $\alpha_1$ and $\alpha_2$ at 3 and 10 TeV muon colliders, where the solid curves include the assumption $\alpha_3=0$ and the dashed ones do not.
\end{itemize}
As shown in Figure~\ref{fig:contoura1a2all}, the 10 TeV muon collider provides a unique sensitivity to the $\bar \mu\mu H$ vertex, significantly better than the 3 TeV option. 
As can be seen, we obtain the following $95\%$ confidence-level bounds at a 3 TeV muon collider
\begin{align}
   & |\Delta \alpha_1| \lesssim 0.75\,,&  &|\alpha_2| \lesssim 0.4&  &{\rm with~no~assumptions~on~}\alpha_3\,, \\
    &|\Delta \alpha_1| \lesssim 0.7\,, & & |\alpha_2| \lesssim 0.4 &  &{\rm assuming~}\alpha_3=0\,,
\end{align}
and at a 10 TeV muon collider
\begin{align}
   & |\Delta \alpha_1| \lesssim 0.1\,,&  &|\alpha_2| \lesssim 0.15&  &{\rm with~no~assumptions~on~}\alpha_3\,, \\
    &|\Delta \alpha_1| \lesssim 0.1\,, & & |\alpha_2| \lesssim 0.1 &  &{\rm assuming~}\alpha_3=0\,.
\end{align}
Also, we notice that these multi-boson production processes provide a chance to determine the sign of the $\alpha_1$, which is not possible in multi-Higgs production and $H\to \mu^+\mu^-$ decay.

    % \begin{figure}[!t]
    %   \centering
    %   \includegraphics[width=0.49\textwidth]{figs/VBha1a2_a30_2machines.pdf}
    % \includegraphics[width=0.49\textwidth]{figs/VBha2a3_a11_2machines.pdf}
    %   \caption{Same as Figure~\ref{fig:contour45V} in the $(\alpha_1,\,\alpha_2)$ plane (left plot),  and in the $(\alpha_2,\,\alpha_3)$ plane in the right plot, respectively. All the Higgs-associated gauge boson production processes that are dependent on $\alpha_3$ at a 3 TeV muon collider (dashed curves) and a 10 TeV muon collider (solid curves) have been combined. The following assumptions are adopted: $\alpha_3=0$ (left), $\alpha_1=1$ (right). For this reason a line corresponding to  $\SMEFTs$ scenario has not been displayed. }
    %   \label{fig:contourVHassumption}
    % \end{figure}

\section{Summary}
\noindent
In this work, we studied the anomalous $\bar{\mu}\mu H^n$ interactions within the EFT framework at a multi-TeV muon collider. We considered all possible processes involving the direct production of electroweak bosons ($W$, $Z$, and $H$) with up to five final-state particles. Our results demonstrate that multi-Higgs production processes offer a unique opportunity to measure the $\bar{\mu}\mu H^n$ vertex. Furthermore, by combining Higgs-associated gauge boson production with multi-gauge boson production processes, a 10 TeV muon collider can precisely constrain the parameters $\alpha_1$ and $\alpha_2$, simultaneously.

\acknowledgments

YM acknowledges support from the COMETA COST Action CA22130.
WK, NK, and TS were supported by the Deutsche Forschungsgemeinschaft
% (DFG, German Research Foundation)
under grant 396021762 - TRR 257.

\bibliographystyle{JHEP}
\bibliography{ref}

\end{document}